\documentclass[aps,twocolumn,showpacs,prb,superscriptaddress]{revtex4-1}
\usepackage{epsf}
\usepackage{graphicx}
\usepackage{tabularx, booktabs, makecell, amsmath,amssymb}
\usepackage{dcolumn}
\usepackage{bm}
\usepackage{color}
\usepackage{mathbbol}
\usepackage{nccmath}
\usepackage[dvipsnames]{xcolor}
\usepackage{graphics}
\usepackage{natbib}
\usepackage{dsfont}
\usepackage{bbold}
\usepackage{appendix}
\def\ga{g_{\alpha}}
\def\ja{j_{\alpha}}
\def\Ja{J_{\alpha}}
\def\Xa{X_{\alpha}}
\def\Wa{Z_{\alpha}}
\def\fa{f_{\alpha}}
\def\za{\zeta_{\alpha}}
\def\phia{\Phi_{\alpha}}
\def\phib{\bar{\Phi}}
\def\psib{\bar{\Psi}}
\def\chib{\bar{\chi}}
\def\sa{\sigma_{\alpha}}
\def\Va{V_{\alpha}}
\def\Ta{T_{\alpha}}
\def\fa{f_{\alpha}}
\def\gama{\gamma_{\alpha}}
\def\Ga{\Gamma_{\alpha}}

\def\Gd{G^{\dagger}}
\def\om{\omega}
\def\Om{\Omega}
\def\gf{\mathcal{G}}
\def\gr{\mathbb{G}}
\def\ar{\mathbb{A}}

\def\myeq{\stackrel{\rm def}{=}}

\def\etaad{\eta_{\alpha}^{\dagger}}
\def\V{\mathcal V}

\begin{document}

\title{Theory of Non-equilibrium Heat transport in  anharmonic multiprobe systems at high temperatures}

\author{Keivan Esfarjani}
\affiliation{Department of Mechanical and Aerospace Engineering, University of Virginia, Charlottesville, Virginia
22904}
\affiliation{Department of Materials Science and Engineering, University of Virginia, Charlottesville, Virginia
22904}
\affiliation{Department of Physics, University of Virginia, Charlottesville, Virginia
22904}
\thanks{k1@virginia.edu}

\date{\today}

\begin{abstract}
We consider the problem of heat transport by vibrational modes (conduction) between Langevin thermostats connected by a central device. The latter is anharmonic and can be subject to large temperature differences and thus be out of equilibrium. We develop a self-consistent Green's function formalism to describe high-temperature and non-equilibrium transport, and derive a formula for the heat current for up to quartic anharmonicity (4th-order in the potential energy). We show the importance of including quartic terms in the anharmonic potential in order to properly describe thermal expansion and temperature dependence to leading order in anharmonicity. This formalism paves the way for accurate and efficient modeling of thermal transport in highly non-equilibrium situations beyond perturbation theory.     
\end{abstract}

\maketitle

%\underline{\sl Introduction}:
\section{ Introduction}

Transport theories of non-interacting quantum systems based on the Keldysh formalism\cite{keldysh}, which treats nonequilibrium flow of charge or heat carriers in a one-dimensional (1D) geometry have been developed in the past. To the best of our knowledge, the first such development was done in 1971 by Caroli et al.\cite{Caroli1971a} where a Green's function formalism was used to describe dynamics of electrons in a 1D crystal. Following the seminal work of Caroli et al, many other groups worked on similar formalisms and proved a formula for the transmission through the system,  now widely used for both non-interacting electrons and phonons. The equilibrium version of it, namely $T={\rm Tr} [ G \Gamma_L G^{\dagger} \Gamma_R ]$ , where $G$ and $\Gamma$ are respectively the retarded Green's function and the escape rates to the leads, was established by Meir and Wingreen\cite{meir-wingreen} and in a similar form by Pastawski\cite{pastawski91} in 1991. This formula holds for a non-interacting (or harmonic, in the case of phonons) system {\it near} equilibrium, meaning the chemical potential or temperature gradients are to be infinitesimally small. These assumptions might not always be realistic, especially in small (mesoscopic) systems subject to temperature differences over fractions of a micrometer, and a formulation for non-equilibrium situations and interacting systems is preferable for the sake of testing the domain of validity of the equilibrium formulas and more accurate description in the case of strong interactions and large driving fields.

\begin{figure}[h!]
\centering
\includegraphics[scale=0.37]{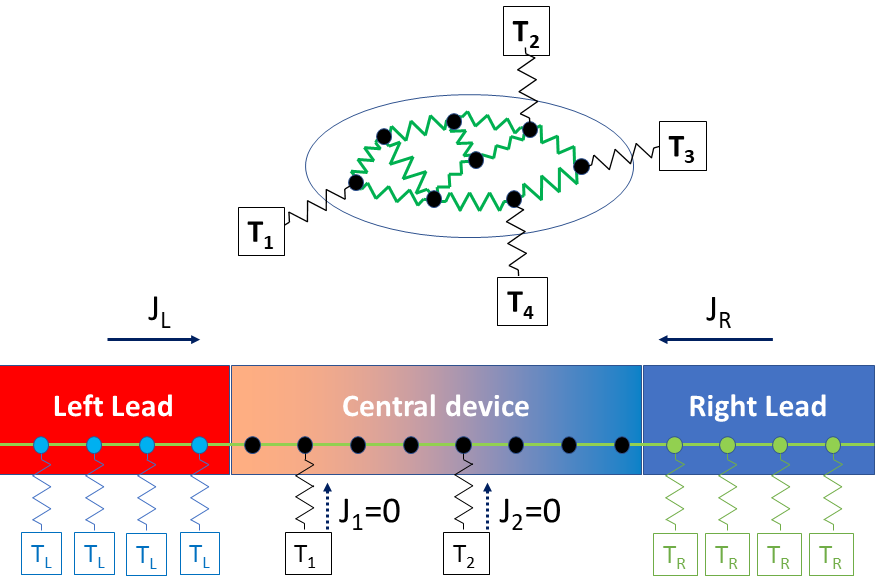}
\caption{Top: a general ``molecular'' multiprobe geometry where the device defined by atoms within the ellipse are connected to 4 reservoirs imposing a temperature or measuring a current. Bottom: the 2-probe geometry. Atoms in each lead are connected to a harmonic thermostat at fixed temperature $T_L$ and $T_R$. In the Buttiker probe, also called the self-consistent reservoirs geometry, to measure the "local" temperature, each layer of the central device may also be weakly connected to a fictitious thermostat at a temperature to be (self-consistently) determined so that the net heat flow from that probe to the device is zero.}
\label{LCR}
\end{figure}

The basic geometry of our problem is a multi-probe one where the system in which scatterings occur is connected to multiple reservoirs or contacts, which impose their chemical potential and/or temperature,  and cause flow of charge or heat carriers (see Fig. \ref{LCR}). This model is used for mesoscopic systems where the carrier mean free path could be on the order of the system length, implying Ohm's law of addition of resistances in series does not necessarily hold, and coherence can play an important role.  The geometry of the reservoirs is  fundamentally one-dimensional (1D), and if there is translational symmetry perpendicular to the current flow, one can use Bloch's theorem to decouple the 3D system into many non-interacting 1D systems, each labeled by a quantum number which is the transverse momentum. So in what follows, we assume such decoupling has been done and we will be dealing with strictly 1D semi-infinite leads, although the central device is arbitrary in shape and structure and maybe connected to multiple 1D probes. 
In this paper, we will be interested in transport of anharmonic phonons, or more generally, vibrational modes,  in mesoscopic systems under large temperature differences. For simplicity, we will use a {\it classical} description. A generalization to the quantum case will be inferred at the end. So in our classical treatment, the frequency $\omega$ is just frequency of vibrational modes and not the energy of phonons. This classical formalism avoids fancier mathematics involving commutation relations, and concepts such as time-ordered or contour-ordered Green's functions. It will only involve "retarded" or causal Green's functions, which help us solve a differential equation in the frequency domain.  Typical considered geometries will be identical to a non-equilibrium molecular dynamics (NEMD) setup where the two ends of the system are attached to two thermostats at different temperatures, and one is interested in measuring the interfacial thermal conductance (see Fig. \ref{LCR}).

%Just like in a NEMD simulation, we follow the dynamics of the system instead of adopting a distribution function approach used in statistical mechanical treatments. %We will also provide a formula for the entropy generation rate. This leap is possible after we perform ensemble averages using the properties of the leads playing the role of a thermostat.

The non-equilibrium anharmonic phonon problem has been addressed in the past by Mingo\cite{Mingo2006} and separately by Wang\cite{jswang2006}  in 2006. They used the many-body perturbation approach of non-equilibrium quantum systems based on the Keldysh formalism, (also called the Non-Equilbrium Green's Function or NEGF method) and derived a lowest-order approximation for the transmission function. Dai and Tian\cite{tian2020} recently implemented this rigorous formulation to calculate the effect of cubic anharmonicity on the phonon transmission function through an ideal interface, applied to Si/Ge and Al/Al with two different masses.
A similar model, which explicitly incorporates transverse momentum dependence, was also developed recently by Guo et. al. \cite{Guo2020}, based on previous work by Luisier\cite{luisier2012}. Polanco has a recent review of these methods based on NEGF\cite{polanco2021}.
These NEGF-based models, although fully quantum mechanical, do not include anharmonicity beyond cubic order nor any thermal expansion effects. 
Other calculations of the transmission in the non-equilibrium regime\cite{Saaskilahti2013} based on the Green's function method, have been based on the self-consistent reservoirs (also called Buttiker probe method,) which was first proposed by Bolsterli et al. in 1970\cite{visscher1970}. In this method, as shown in Fig. \ref{LCR}, every layer is connected, with a {\it very weak} coupling, to a fictitious probe at a given temperature with which it becomes into equilibrium. The probe temperature, which is also the assigned temperature of that layer, is obtained from the constraint that the net heat current from the system to the added fictitious probe should be zero. Note that the system could be out of equilibrium, so that a temperature is not really well-defined at a given layer. This shortcoming is also present in NEMD simulations where the assigned "temperature" of a layer is just the average kinetic energy of atoms in that layer, although there is no evidence of local thermal equilibrium. 
So even though non-equilibrium effects are included through this Buttiker probe method, it does not include anharmonicity. We should mention at this point that there is numerical evidence of absence of equipartition in the vibrational modes near the interface\cite{nam2017,feng2019}, implying that a definition of local temperature is not really justified near an interface.
In another work, based on the MD simulation, which fully includes anharmonicity, Saaskilahti et al.\cite{Saaskilahti2014} extended the harmonic formulation of the transmission function based on the actual MD trajectories by Chalopin et al.\cite{chalopin2012a,chalopin2013}, to include anharmonic corrections.
In the actual calculation, arguing that the anharmonic part of the current is usually small, they only used its harmonic formula but with velocities and positions coming from the full anharmonic atomic trajectories, in order to deduce the interfacial thermal conductance. The advantage of this approach over NEMD is that the heat current and thermal conductance can be decomposed in the frequency domain.  
Approaches based on MD trajectories, while including full anharmonicity, suffer from noise and would require a large number of simulations in order to perform proper ensemble averaging, whereas many-body approaches might be inaccurate if a perturbative expansion in powers of anharmonicity is used, but otherwise do not suffer from noise and treat ensemble averaging analytically.
In this work, we try to overcome these limitations by adopting a non-perturbative  many-body  approach by fully including in the current the effect of anharmonic terms introduced in the Hamiltonian, without using Buttiker probes. Furthermore, using a classical method to derive an expression for the heat current, we argue that to leading order, it is necessary to include quartic terms in the Hamiltonian, in order to properly describe both the thermal expansion and the dominant temperature dependence effect in the heat current.

%\underline{\sl Dynamics}:
\section{Dynamics}

We start by defining our model and the assumptions. A multiprobe geometry is assumed as shown in Fig. \ref{LCR}  in which a central region, also called the ``device" is connected to many semi-infinite one-dimensional (1D) leads, playing the role of thermostats imposing a temperature at the boundaries of the system. We will not be concerned with temperature drops in the thermostats, which are assumed to be harmonic and follow Langevin dynamics. 
If needed, parts of the leads can be incorporated in the device region to illustrate the temperature profile.
The Hamiltonian of the device (D) and its coupling to the lead $\alpha$ are:
\begin{flalign}
H_D &=\sum_{i \in D}  \frac{p_i^2}{2 m_i} + \sum_{ij \in D} \,\frac{1}{2!} \phi_{ij}\, u_i\, u_j \\
&+ \sum_{ijk \in D} \,\frac{1}{3!} \psi_{ijk}\, u_i \,u_j \,u_k + ...  \\
%H_{\alpha} &=\sum_{l \in \alpha} \Big( \frac{p_l^2}{2 m_{\alpha}}  +\gama u_l {\dot{u}}_l - \frac{\zeta_l u_l}{\sqrt m_{\alpha}}+ \sum_{l' \in \alpha}\, \frac{\phi_{\alpha,ll'}}{2!}  \, u_{l}\, u_{l'} \Big) \label{left} \\
H_{\alpha,D} &=  \sum_{i \in D} \sum_{l \in \alpha} \,
W_{\alpha,il} \, u_i \,u_l    \label{LD}
\end{flalign}
%where  $m_\alpha$ is the mass of every atom in the lead $\alpha$. 
The dynamical variable $u_i(t)$ refers to the displacement of atom $i$ about the zero-temperature equilibrium position (the force on atom is zero for $u=0$).
The leads labeled by $\alpha$ are semi-infinite chains following Langevin dynamics. 
%As will be explained in section \ref{lin_anh}, part of the anharmonic terms maybe assumed to be reabsorbed in the harmonic terms leading to an effective Hamiltonian.
After the standard change of variables to $x_i(t)=\sqrt{m_i} u_i(t) $, %where $u_i(t)$ is the atomic displacement of atom $i$ at time $t$, 
and to 

$$\Phi_{ij}= \frac{\phi_{ij}}{\sqrt{m_i m_j}}= \frac{1}{\sqrt{m_i m_j}} \frac{\partial^2 H_D}{\partial u_i \partial u_j }  (u=0)$$  

$V_{\alpha,il}= W_{\alpha,il}/\sqrt{m_i m_\alpha}$ and $\Phi_{\alpha,ll'}= \phi_{\alpha,ll'}/m_\alpha$ we arrive at the following equation of motion for atoms in the central region:
\begin{equation}
    \frac{d^2x}{dt^2} = -\Phi\, x - \sum_{\alpha} V_{\alpha}\, x_{\alpha} + a
\label{eom_center}
\end{equation}
Note we have used capitalized Greek letters ($\Phi, \Psi,...$) for mass-rescaled force constants, and lower-case Greek letters  ($\phi, \psi,...$) for the bare potential energy derivatives. The letter $\alpha$ refers to the leads, and the  % and therefore the following equations remain valid for a multiprobe configuration.
dynamical variable $x=(x_1,...,x_N)$ can be thought of as an array containing the displacements of all the atoms in the central region (also called device),  $\Phi$ as the force constant matrix between such atoms, and $V_{\alpha}$ as the force constant matrix connecting atoms of the lead $\alpha$ to atoms in the device. Finally, $a=-1/2 \Psi  x^2+...$ is the anharmonic part of the force, which, for now, we keep as $a$ for brevity. The dynamics of atoms in the lead is of {\it Langevin} type, where a set of identical coupled harmonic oscillators are subject to  damping $\gama$ and noise $\za$ as follows:
\begin{equation}
    \frac{d^2x_{\alpha}}{dt^2} = -\Phi_{\alpha} x_{\alpha} - V_{\alpha}^T x - \gamma_{\alpha} \frac{dx_{\alpha}}{dt} + \zeta_{\alpha}
\label{eom_leadt}
\end{equation}
where the superscript $T$ stands for transpose.
The force constants $ Phi_{\alpha}$ can be thought of as effective FCs at the temperature of interest, so that we do not need to introduce anharmonicity in the leads, which merely play the role of absorbing phonons from the device and reinjecting thermalized phonons into the device.
We will proceed by eliminating the lead variables $x_{\alpha}$ in Eq. \ref{eom_center} using the Green's function method. To this end, we start by taking the Fourier transform of the above two equations according to:
\begin{equation}
X(\omega)=\int dt \,x(t)\, e^{i \omega t} ; \,\, x(t)=\int \,\frac{d\omega}{2 \pi} \,X(\omega)\, e^{-i \omega t}    
\end{equation}
The equations of motion become:
\begin{flalign}\label{eom-device}
-\omega^2 X  &= -\Phi X - \Sigma_{\alpha} V_{\alpha}\, X_{\alpha} + A \\   
-\omega^2 X_{\alpha}  &= -\Phi_{\alpha} X_{\alpha} - V_{\alpha}^T X + i\omega \gamma_{\alpha} X_{\alpha} + \zeta_{\alpha}(\omega)  
\label{eom_leadw}
\end{flalign}
Note that the frequency-domain variables are represented with capitalized letters.
Now that the differential equations are transformed to algebraic ones, one can easily proceed to eliminate the lead degrees of freedom in the main equation of motion by using the green's functions.
Let  $\ga(\omega)$ be the retarded (causal) Green's function associated with the lead $\alpha$. The positivity of the damping factor $\gama$ insures causality. The solution to Eq. \ref{eom_leadw} after the transients have decayed to zero, can be written as:
\begin{equation}
    \Xa  = \ga(\omega)\, (\za -\Va^T X) 
    \label{xalpha}
\end{equation}
where 
\begin{equation}
\ga^{-1}=[-\omega^2 -i \omega \gama + \phia] 
\label{sgf}    
\end{equation}
In Eq. \ref{eom-device} we need $-\Va \Xa =  \eta_{\alpha} + \sigma_{\alpha} X$, which we obtain from Eq. \ref{xalpha}. Here $\eta_{\alpha}(\omega)=-\Va \ga(\omega) \za(\omega) $, and $\sa= \Va \, \ga \Va^T$.
Likewise defining the retarded Green's function of the central region as  $G^{-1}=[-\omega^2  + \Phi -\sum_{\alpha} \sigma_{\alpha}(\om) ] $, we can write the solution to the central region as:
\begin{equation}
 X (\omega)= G(\omega)\, \Big[\sum_{\alpha} \eta_{\alpha}(\omega)\, +A(\omega) \Big] 
    \label{dispw}
\end{equation}

The function $\sa(\omega)= \Va \, \ga(\omega) \Va^T$ is traditionally called the self-energy of lead $\alpha$, and shows the effect of this lead on the spectrum of  the device which is given by the poles of $G$. Its real part provides a correction to the eigenvalue spectrum $\om^2$, and its imaginary part, divided by $2 \omega$, gives the inverse lifetime of an excitation of the central region caused by interactions with the lead. It is the rate at which the excitation leaks into the lead. Omitting the transient contribution of the initial conditions, this is the solution to the equations of motion, which depend on the stochastic functions $\eta_{\alpha}=-V_{\alpha} g_{\alpha} \za$.  

%\underline{\sl Physical observables}:
\section{Physical observables}

 The equations of motion we derived are deterministic for every realization of the random forces. To simulate the real thermodynamical behavior of the baths, so that a temperature can be assigned to them, we need to perform an ``ensemble'' average, denoted by $\langle \hdots \rangle$ over all realizations of the forces subject to the constraint imposed by the fluctuation-dissipation (FD) theorem: 
% {\setlength{\abovedisplayskip}{0pt}%
$$
  \langle \zeta_{i,\alpha}(t) \rangle = 0 
$$
  \begin{equation}
%  \boxed{ 
 \langle \zeta_{i,\alpha}(t) \zeta_{j,\alpha'}(t') \rangle 
 %&\stackrel{FD}{=}
 =2 \gama k_B T_{\alpha} \,\delta(t-t')\delta_{\alpha,\alpha'}\delta_{ij}
% }
  \label{fd}
 \end{equation}

The noise is white and different sites $i,j$ of leads $\alpha,\alpha'$ are uncorrelated with each other. Physical observables are then obtained after an ensemble average is performed over forces. This is where irreversibility is introduced in this deterministic formalism, as a result of which, entropy is generated in the device. The latter can be understood as the log of the distribution function of  the device as the random forces are varied with the constraints imposed by the FD theorem.. 

The main quantity of interest is the heat current. The heat from lead $\alpha$ can be defined as the net rate at which energy is flowing into the device from that lead. It is the work done from lead $\alpha$ on the device's degrees of freedom per unit time, which is the product of the velocity degrees of freedom of the device times the force from lead acting on them: $j_{\alpha} (t)= {\rm Tr} [\dot{x} \, (-\Va x_{\alpha})^T ]$, where the trace is taken over the device degrees of freedom after using Eq. \ref{xalpha}. Since the current depends on the stochastic functions $\zeta$, we will take its ensemble  
average to find the response of the system to an applied temperature difference. We will start by taking the Fourier transform of $j_{\alpha} (t)$ and call it $\Ja (\Omega)$. Note due to time integration, we must have $\langle  \Ja(\Om=0) \rangle=\tau \langle  j_{\alpha} \rangle$, where $\tau \to\infty$ is the time integration window: 

\begin{widetext}
\begin{equation}
\langle  \Ja(\Om) \rangle  = \int \frac{d\omega}{2 \pi} \,  \omega \,  {\rm Tr} \,\Big[ \Im \langle X(\omega) \eta_{\alpha}^{\dagger}(\omega-\Om) \rangle  -  \langle X(\omega)   X^{\dagger}(\omega-\Om) \rangle \,\Ga(\omega-\Om)/2 \Big]
\label{heatcurrent}
\end{equation}
\end{widetext}

where we used  the notation $\Gamma_{\alpha} =-i(\sa-\sa^{\dagger})=2 \Im (\sa) $ for twice the imaginary part of the lead $\alpha$ self-energy. 
The DC response is found by taking the $\Om \to 0 $ limit.
Note that because we are interested in the DC response, only diagonal terms of correlations ($\Om=0$) are needed here.  
Thus the calculation of the heat current is reduced to the calculation of the two correlation functions and the so-called lead self-energy $\sa(\omega)$, followed by a frequency integration.
Note that this current from lead $\alpha$ is the sum of two terms: the first one, proportional to $\Wa=\langle X(\omega) \, \eta_{\alpha}^{\dagger}(\omega) \rangle$, is the work per unit time of the stochastic forces on the device,  while the second one $\langle X(\omega) \, X^{\dagger}(\omega) \rangle \Ga^{\dagger}$, proportional to a displacement autocorrelation, is the work of the lead dampers trying to reabsorb some of the excess energy injected from the device in order to re-establish thermal equilibrium in the lead. 
 
The average denoted by $\langle  \rangle$ is an average over the stochastic forces in the thermostats which have white noise characteristics. When it is performed over the device degrees of freedom, the leads being at different temperatures, it becomes a {\it non-equilibrium} average and can only be calculated using the equation of motion \ref{dispw} and the statistical properties of the Langevin thermostats, i.e. the fluctuation-dissipation theorem. For anharmonic interactions involving higher powers of displacements in $A$, the calculation of current will lead to a hierarchy of equations, each containing higher powers of displacements, and has so far been computed using different approximations\cite{Mingo2006,jswang2006,Tian2014a}. Another quantity of interest is the entropy generation rate in the device which can be expressed as: $\dot{S}=-\sum_{\alpha} \langle  j_{\alpha} \rangle /T_\alpha $

%\underline{\sl Constraints}:
\section{Constraints}

Before proceeding to the calculation of the correlation functions, we recall two constraints that the heat current needs to satisfy. The first is the {\it  detailed balance} relation which states that if all leads are at the same temperature $T$, the net current $\langle  j_{\alpha} \rangle$ %and even each $\omega$ component in the frequency integral  
should be identically zero for all leads $\alpha $. %:  $\, \Ja(T_{L}=T_R=T)=0$. 
The second constraint is that of {\it current conservation}, which in steady state $\Om \to 0$, and under no additional heat generation in the device, reduces to $\Sigma_{\alpha} \langle j_{\alpha} \rangle =0$. This is also known as the {\it Kirchhoff's law} in the context of electrical circuits. Note that AC components of the current need not satisfy this constraint as they reflect the information on transient currents during the relaxation process, and depend on the heat capacity of the system. 
Any physically correct description of transport, should exactly satisfy these two constraints.
 
%\section{Effective harmonic force constants: Linearization of the anharmonic terms} \label{lin_anh}
%\underline{\sl Thermal expansion}:
\section{Thermal expansion}

As the temperature of a system is raised, there can be thermal expansion due anharmonicity. The equilibrium position of the atoms is shifted, and this will also cause a change in the force constants as bond lengths have changed. To take these effects into account, while simplifying the notations, we will slightly modify the formalism as follows:
the displacement variable $X$ is changed to $Y(\om)=X(\om)-\langle X \rangle$ or $y(t)=x(t)-\langle x \rangle$ which has zero average by construction. The resulting nonlinear equations satisfied by $\langle X \rangle $ are derived by taking the average of the equation of motion \ref{dispw} or equivalently setting the average force on each atom to zero (see appendix \ref{xtoy} for more details).
\begin{equation}
    \langle \partial \V / \partial x \rangle = \Phi \langle x \rangle+ \Psi \langle x x \rangle /2 +...=0
    \label{favg0}
\end{equation}
The resulting equations will depend on correlations such as $\langle  Y_i Y_j \rangle $, $\langle Y_i Y_j Y_k \rangle $ and higher powers.  
Accordingly, the potential energy derivatives will be evaluated at the zero of $Y$, and will be denoted with a bar sign on top of them ($\Phi \to \phib$ etc...).
While the variable $X$ satisfies the equation of motion: $$-\omega^2 X = -\partial \V /\partial X -\sum_{\alpha} V_{\alpha} X_{\alpha}\,\, ,$$ the new variable $Y$ satisfies:
 $$-\omega^2 Y = -\partial \V /\partial X +\langle \partial \V /\partial X \rangle -\sum_{\alpha} V_{\alpha} X_{\alpha}$$

Next we will linearize the forces with respect to $y$:
\begin{flalign}
-\frac{\partial \V}{\partial y}=- \big(\frac{\partial \V}{\partial y}\big)_{y=0} - \big(\frac{\partial^2 \V}{\partial y^2}\big)_{y=0} \, y + a(y) = -\phib y + a(y) \nonumber 
\end{flalign} 
where we have set $(\frac{\partial \V}{\partial y})_{y=0} = 0$ to define the thermal expansion $\langle x \rangle=x_0$ (see appendix \ref{xtoy}).
As we will show, the effect of temperature will be to renormalize the FCs, not only through thermal expansion but also due the thermal fluctuations as we will show using the non-equilibrium mean-field approximation (NEMF) also detailed in section \ref{nemf}.  
Next, we define a renormalized Green's function using renormalized force constants $\phib=\big(\frac{\partial^2 \V}{\partial y^2}\big)_{y=0}$  as 
\begin{equation}
    \gf^{-1}=[ -\om^2+\phib-\sum_{\alpha} \sa]
    \label{gfbar}
\end{equation}
%The harmonic force constant is renormalized because of the bond length change, according to: $\phib_{ij}=(\partial^2 \V/\partial y_i \partial y_j) _{y=0} =\Phi_{ij}+ \Psi_{ijk}\, \langle x_k \rangle+ \frac{1}{2} \chi_{ijkl}\, \langle x_k \rangle \langle x_l \rangle $.
% captures a major part of the anharmonicity, and has the advantage that the remaining anharmonic part of the forces, $a$, will be small. 
With this GF, the displacements $Y$ satisfy 
\begin{equation}
%\boxed{    
Y=\gf (\sum_{\alpha} \eta_{\alpha}+A) 
%}
    \label{eomy}
\end{equation}
Note one can add any constant $\lambda$ to the force constant  $\phib$ in the above GF, provided $\lambda Y$ is also added to the anharmonic force $A$. We will make use of this freedom in the next section to further simplify the formalism. 

%In terms of $Y$ the heat current is now:
%\begin{flalign*}
%\langle  \ja \rangle  = \int \frac{d\omega}{2 \pi \tau} \,  \omega \,  {\rm Tr} \,\Big[ \Im \langle Y \eta_{\alpha}^{\dagger} \rangle 
% - \big( \langle X \rangle \langle X^{\dagger} \rangle + \langle Y   Y^{\dagger} \rangle \big) \,\frac{\Ga}{2} \Big]
%\label{heatcurrentY}
%\end{flalign*}

%{\underline  {\sl Force constant Renormalization}} 
\section{Force constant Renormalization}
\label{nemf}
%As we previously mentioned, it is always possible to add a term $\lambda Y$ to the anharmonic force $A$ and add the coefficient $\lambda$ to $1/\gf$ without changing the solution $Y$.  
Given the form of the above equations, we can add $-Y \langle  \frac{\partial A}{\partial Y} \rangle $ to $A$ and add $- \langle \frac{\partial A}{\partial Y} \rangle $ to $1/\gf$ so that now the renormalized GF becomes:
\begin{equation}
    \mathbb{ G}^{-1}=\gf^{-1}-\langle  \frac{\partial A}{\partial Y} \rangle=[-\omega^2+\phib -\langle  \frac{\partial A}{\partial Y} \rangle + \sum_{\alpha} \sigma_{\alpha}]
\end{equation}
while the renormalized anharmonic force now becomes $\mathbb {A}=A-Y\langle  \frac{\partial A}{\partial Y} \rangle$ in the right hand side of  Eq. \ref{eomy}. This renormalization of harmonic force constants captures a major part of anharmonicity (because $\langle \partial \ar /\partial Y \rangle=0$), and is in spirit very similar to the lowest-order self-consistent phonon theory, which in the past has been applied to equilibrium systems. The advantage of this renormalization is that, as we will see, the lowest anharmonic correction in $\Wa$  disappears by construction since $\langle \partial \ar /\partial Y \rangle =0$, leading to the smallest variance and higher moments of $\langle \partial \ar /\partial Y \rangle $.

\begin{figure}[!h]
\centering
\includegraphics[scale=0.40]{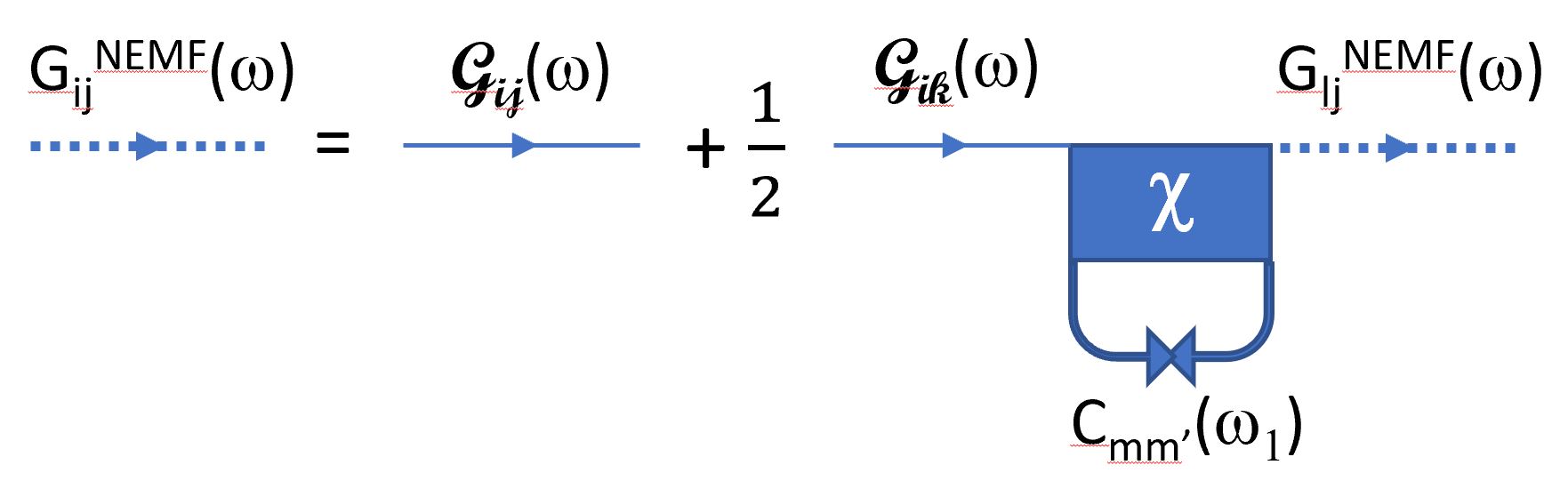}
\caption{Feynman diagram associated with $\gr$ represented by thick dashed lines.  The phonon Green's function $\gf$ is shown with thin solid lines, and the quartic vertex $\chi$ with the solid square. The thick line with opposite arrows represents the displacement autocorrelation  $C(\om_1)$. Internal frequency $\om_1$ is integrated over. 
}
\label{gfeynman}
\end{figure}

We will refer to this choice of the reference GF as the {\it Non-equilibrium mean-field} approximation (NEMF).
With this choice, the equation of motion for $Y$ becomes:
\begin{equation}
\boxed{    Y=\gr (\sum_{\alpha} \eta_{\alpha}+\ar) }
    \label{reom}
\end{equation}
The Feynman diagram associated with the new Green's function $\gr$ is shown in Fig. \ref{gfeynman}.

%\underline{\sl Displacement-noise
\section{Displacement-noise correlations}

One can see from Eq. \ref{heatcurrent} that the calculation of the heat current requires the calculation of the displacement-noise correlation $\Wa$ and displacement autocorrelations $C$. We proceed to the calculation of these quantities first within the harmonic approximation, and then in the presence of anharmonic forces of the form $a =- \psi \, y^2/2- \chi \, y^3/6$. %Recall that the ``averages'' in the correlation functions are evaluated in the central region and are therefore {\it non-equilibrium} averages. 

Let us start with the noise autocorrelation which will appear in the calculation of displacement-noise correlation $\Wa$. In the frequency domain, using the fluctuation-dissipation theorem Eq. \ref{fd}, one  can derive (see appendix \ref{noiseFD}):

\begin{flalign}
\boxed{ \langle \eta_{\alpha}(\om) \eta_{\alpha}^{\dagger}(\om) \rangle = \Ga(\om) \frac{k_BT_{\alpha}}{\omega} \tau = \Ga(\om) f_{\alpha} \tau }
%=2 \Im(\sa(\om))  \frac{k_BT_{\alpha}}{\omega} \tau   
%\langle &\eta_{\alpha}(\omega) \eta_{\alpha'}^T(\omega) \rangle  = \Gamma_{\alpha}  \, \frac{k_BT_{\alpha}}{\omega}\, \delta_{\alpha,\alpha'}
 \label{noiseautocorrelation}
\end{flalign}
 where, for brevity, we have replaced the ``occupation factor''\cite{hbar} $k_BT_{\alpha}/\omega$ by $\fa$, and $\tau$ represents the integration time which goes to infinity and cancels the $\tau$ in the expression for the current $\ja=J_{\alpha}/\tau$. In the case of a white noise, we show in appendix \ref{noiseFD} that the result will not depend on the thermostat damping parameter $\gamma$, and thus we adopt this type of noise for the thermostats. 
  
 Next, we need to calculate displacement-noise correlations: $\Wa=\langle Y \eta_{\alpha}^{\dagger} \rangle $. 
Since  $-i\omega Y$ is a velocity, $\omega \Im (\Wa)$ is the power exerted by the random force $\eta$ on the device and therefore can be interpreted as the  {\it heat injected per unit time and unit frequency (mode) from lead $\alpha$ into the device}. 

In the harmonic case ($A=\ar=0; G=\gf=\gr$), using the equation motion \ref{reom}, this expression is simplified to:
\begin{equation}
 \Wa^H(\omega)=\gf \langle \eta_{\alpha} \,\eta_{\alpha}^{\dagger} \rangle=\gf \Ga \fa \tau 
\end{equation} 
For non-zero anharmonicity, the three GFs are different, and adopting $\gr$ as the reference GF, we have the NEMF approximation to the displacement-noise correlation function as: 
\begin{equation}
 \Wa^{NEMF}(\omega)=\gr \langle \eta_{\alpha} \,\eta_{\alpha}^{\dagger} \rangle=\gr \Ga \fa \tau 
\end{equation}

To go one step further and include the effect of anharmonicity in $\Wa$,  we will use the Novikov-Furutsu-Donsker identity\cite{tabar2019analysis} (for a proof also see appendix\ref{nfdapp}), which states that for any functional of the white noise $f[\eta]$, we have 
\begin{equation}
\boxed{
\langle f[\eta] \, \eta_{\alpha}^{\dagger}(\om) \rangle  =   \langle \frac{\delta f[\eta]}{\delta \eta_{\alpha}} \rangle \langle \eta_{\alpha} \, \eta_{\alpha}^{\dagger} \rangle 
}
\label{NFD}
\end{equation}
It has the advantage of lowering the powers of $\eta$ in $f$. Using this theorem, we have: 
 $ \langle Y \eta_{\alpha}^{\dagger} \rangle= 
 \langle \frac{\partial Y}{\partial \eta_{\alpha}} \rangle \langle \eta_{\alpha} \, \eta_{\alpha}^{\dagger} \rangle $.

From the equation of motion Eq. \ref{reom} and the chain rule, we find 
\begin{flalign}
 % \langle \frac{\delta A[\eta]}{\delta \eta_{\alpha}} \rangle &= \langle \frac{\partial A}{\partial Y}    \frac{\partial Y}{\partial \eta_{\alpha}} \rangle ;\, % \nonumber \\
   \frac{\partial Y}{\partial \eta_{\alpha}}  =  (1-\gr \frac{\partial \ar}{\partial Y} )^{-1}  \gr=\gr + \gr \frac{\partial \ar}{\partial Y}\gr+\gr \frac{\partial \ar}{\partial Y}\gr \frac{\partial \ar}{\partial Y} \gr+...  \nonumber
\end{flalign}
so that finally,

\begin{widetext}
\begin{equation}
%\boxed{ 
\Wa= \langle Y \eta_{\alpha}^{\dagger} \rangle = \langle \frac{\partial Y}{\partial \eta_{\alpha}} \rangle \langle \eta_{\alpha} \, \eta_{\alpha}^{\dagger} \rangle= \langle  
\big(1-\gr \frac{\partial \ar}{\partial Y} \big)^{-1}  \rangle \, \Wa^{NEMF} =
\big(1+\gr \langle \frac{\partial \ar}{\partial Y} \gr \frac{\partial \ar}{\partial Y} \rangle + \gr \langle \frac{\partial \ar}{\partial Y} \gr \frac{\partial \ar}{\partial Y} \gr \frac{\partial \ar}{\partial Y} \rangle + ... \big)  \, \Wa^{NEMF}   
\label{dyson}
\end{equation}
\end{widetext}

Note that, by construction,
the first term involving $\langle \partial \ar / \partial Y \rangle$ is identically zero.

In the language of many-body theory, this is the expanded form of Dyson's equation  involving thermal average of powers of the anharmonic force derivatives. This exact result  cannot be calculated, because the average of the inverse of powers of $Y$ cannot be calculated exactly, and instead one may add the power series term by term provided the sum is convergent.

This is one of the main results of this paper, providing a more accurate and likely convergent expression for the displacement-noise correlations, and leading to a simple calculation of heat currents if cubic terms are to be neglected (using the NEMF reference Green's function $\gr$). 

The next-order correction consists in adding the effect of anharmonicity to second order, i.e. writing the displacement-noise correlation function as:
\begin{equation}
\Wa \approx 
\big(1+ \gr \langle \frac{\partial \ar}{\partial Y} \gr \frac{\partial \ar}{\partial Y} \rangle  \big)  \, \Wa^{NEMF}  
%} 
\label{2nd_order_z}
\end{equation}
with $ \ar =A-Y \langle \frac{\partial A}{\partial Y} \rangle=-\frac{1}{2}\psib Y Y -\frac{1}{6} \chi (YYY-3Y\langle YY \rangle)$
and 
$\frac{\partial \ar}{\partial Y} =\frac{\partial A}{\partial Y} -\langle \frac{\partial A}{\partial Y} \rangle=-\psib Y -\frac{1}{2} \chi (YY-\langle YY \rangle)
$. We can note that the major part of the quartic anharmonicity has been removed, and the expansion is in powers of $\partial \ar /\partial Y$ which is centered at zero and has therefore the smallest higher moments. 
When raised to second power in the above formula for $\Wa$, cubic and quartic terms become decoupled if we neglect averages of terms of odd power in $Y$ which are expected to be small if non-zero. At low-temperatures or weak anharmonicity, the dominant contribution to the second-order terms can be written as:
\begin{equation}
\Wa \approx 
\big(1+  \langle \gr \psib Y \gr \psib Y  \rangle  \big)  \, \Wa^{NEMF}  
\label{zcubicdominant}
\end{equation}
An explicit form of this equation is provided in the later section and in appendix Eq. \ref{zeqs}.
 
%    \begin{figure}[h]
%    \centering
%    \includegraphics[scale=0.55]{dyson.png}
%    \caption{Feynman diagrams associated with the approximated Dyson equation \ref{apdyson} for the displacement-noise correlations $\Wa(\om)$. Letters i,j,k,l,m refer to device degrees of freedom (atom, cartesian coordinate), dotted and solid lines represent respectively the  GF and the function $\Wa$, the small circle represents $\Ga$ and the square with $D$ represents $\gf \langle \frac{\partial A}{\partial Y} \rangle$, which to the first-order reduces to the second diagram involving only the quartic term $\chi$ and the displacement autocorrelation $C(\om_1)$. There is an implicit integration over the internal frequency $\om_1$. To second-order, one can add the cubic term and obtain the thrid diagram for $D$. }
%    \label{zdiagram}
%    \end{figure}

%\underline{\sl Displacement autocorrelations}:
\section{Displacement autocorrelations}

Finally, the last correlation function needed is $C(\om)=\langle YY^{\dagger} \rangle$.  Note the autocorrelation matrix $C$ is Hermitian.
The function $\om C(\om)$ can be interpreted as the (non-equilibrium) number of excitations of frequency $\om$ present in the device due to its contact with the leads. If all lead temperatures are equal, we recover the equilibrium occupation times the total DOS, $G (\Sigma_{\alpha}\Ga) G^{\dagger} =2\Im(G) $, which is the total (equilibrium) number of excitations in the device. In the heat current, this term appears as $-\omega C \Ga/2 $ and can be interpreted as the heat current going from (because of the negative sign) the device into this lead as $\Ga/2\om$ is the escape rate into the lead $\alpha$. 

Similar to the treatment of $\Wa$, we will start with the equation of motion, Eq. \ref{reom},
 and the FD theorem, Eq. \ref{noiseautocorrelation}, to write $C$ as:
\begin{flalign}
&C(\omega) = \langle Y(\omega) Y^{\dagger}(\omega) \rangle =
\nonumber 
\\
&\gr(\omega) \Big[ \Sigma_{\alpha} \big( \langle \eta_{\alpha}\eta^{\dagger}_{\alpha} \rangle  +  
%\Sigma_{\alpha} 
\langle  \ar \eta^{\dagger}_{\alpha} \rangle + \langle  \eta_{\alpha}  \ar^{\dagger} \rangle \big) +\langle \ar \ar^{\dagger} \rangle \Big] \gr^{\dagger}(\omega) 
\label{ctot}        
\end{flalign}

The first term is the NEMF contribution and is reduced to a known result, with now the renormalized GF $\gr$ being used instead of the standard harmonic one ($G$ or $\gf$), in order to include thermal effects to some extent:  
\begin{equation}
    C^{NEMF}(\omega)=\gr\,  \Sigma_{\alpha}\langle \eta_{\alpha}\eta^{\dagger}_{\alpha} \rangle\, \gr^{\dagger}=\Sigma_{\alpha} \, (\gr \Ga \gr^{\dagger}) \,\fa \tau 
\label{CH}
\end{equation}
The NEMF approximation, which consists in using $\gr$ and neglecting the contributions of anharmonicity included in $\ar$,  is very similar to the harmonic approximation. 
Within this approximation, where $C\approx C^{NEMF}$ and $\Wa \approx \Wa^{NEMF}$, the transmission becomes  ${\rm Tr} \,[ \gr \Gamma_L \gr^{\dagger} \Gamma_R]$, very similar to the harmonic result (see appendix \ref{harmonic}), but with the GF substituted by the renormalized $\gr$, and includes the effect of (quartic) anharmonicity to lowest-order.

%It can be shown that the next-order terms in Eq. \ref{ctot} are:
%\begin{flalign}
%        C^{1}(\omega)&= \gr\, (\Wa-\Wa^{NEMF})^{\dagger} + (\Wa-\Wa^{NEMF})  \gr^{\dagger} \\
%        C^{2}(\omega)&= \gr\, \langle   \ar \ar^{\dagger} \rangle   \gr^{\dagger} 
%\label{C1}
%\end{flalign}

Linear terms in $\ar$ lead to terms similar to $\Wa$ which has already been discussed. The only remaining difficulty is with the $\langle \ar \ar^{\dagger} \rangle$  terms which do not explicitly contain any noise term, but have higher powers of displacements. Such terms can only be calculated approximately as the use of the equations of motion will involve higher powers of displacements. To have a second-order approximation consistent with that used for $\Wa$ in Eq. \ref{zcubicdominant}, we have to use:
$$
\langle   \ar \ar^{\dagger} \rangle^{(3)} = C^{(3)}= \frac{1}{4} \psib^2 \sum_{\om_1} C(\om-\om_1) C(\om_1) 
$$

%To be consistent, we will consider adding terms in $\Wa$ and $C$ that contain the same power of temperature or occupation factors $f_{\alpha}$. The quasi-harmonic approximation already contains the terms linear in the occupations, although $\gr$ is also implicitly temperature-dependent. The next correction terms would have the second power of $f_{\alpha}$: in $\Wa$ it will be terms coming from 

It can be shown 
that this approximation, taken to a self-consistent level satisfies current conservation, meaning $\sum_{\alpha} j_{\alpha}=0$, however the spectral components of the current:  $\Im{\Wa}(\om) - C(\om) \Gamma_{\alpha}(\om)/2$ do not necessarily lead to zero when summed over all leads. 
Including for completeness both the cubic and quartic components of the anharmonic forces to second order, the self-consistent set of equations to be solved with the reference Green's function $\gr$  are:
\begin{flalign}
& \Wa = \Wa^{NEMF} + \gr (\Sigma^{(3)}+\Sigma^{(4)}) \Wa \nonumber \\ 
&\Sigma^{(3)}(\om) = 2 \psib^2 \sum_{\om_1} \gr(\om-\om_1) C(\om_1) \nonumber \\
&\Sigma^{(4)}(\om) = \frac{3}{2} \chib^2 \sum_{\om_1,\om_2} \gr(\om-\om_1-\om_2) C(\om_1)C(\om_2) \nonumber \\
&  C=\sum_{\alpha} (\delta\Wa \, \gr^{\dagger} + \gr\, \delta\Wa^{\dagger}) - C^{NEMF} + \gr P \gr^{\dagger} \nonumber \\ 
& \delta\Wa=\Wa-\Wa^{NEMF} \nonumber\\
& P= C^{(3)}+C^{(4)} \nonumber\\
& C^{(3)}= \frac{1}{2} \psib^2 \sum_{\om_1} C(\om-\om_1) C(\om_1) \nonumber \\
& C^{(4)}= \frac{1}{6} \chib^2 \sum_{\om_1,\om_2} C(\om-\om_1-\om_2) C(\om_1)C(\om_2)   
  \label{NESCP}
\end{flalign}
The equations for $\Wa$ and $C$ can also be represented using Feynman diagrams as shown in Figs. \ref{zfeynman} and \ref{cfeynman} in appendices  \ref{zalpha} and \ref{yy}. More explicit forms of these equations are also reproduced in this appendix as Eqs. \ref{zeqs} and \ref{ceqs}.

Starting inputs for $C$ and $\Wa$ could be their NEMF values in the right-hand sides of the above equations, and the latter can be solved iteratively until convergent.  
Note that while the term $\Wa$ requires $\partial \ar/ \partial Y$, the terms $C$ require $\ar$ itself, but these equations contain only second powers of $\ar$ and $\partial \ar/\partial Y$. Once iterations converge, the obtained $C$ and $\Wa$ functions can then be inserted in Eq \ref{heatcurrent2} to compute the heat currents from each lead.

These equations would be the same as  the ones obtained from the many-body non-equilibrium Keldysh formalism with the difference that the ``occupation factors'' $f_{\alpha}=k_B T/\omega$ are classical ones, instead of Bose-Einstein functions. In this sense, they can directly be compared to results from classical non-equilibrium MD simulations, which are exact in anharmonicity but have inherent statistical noise in them.

Another interesting feature to note is the increase of the overall conductance with the temperature (if $\Delta T$  is held small). This is in agreement with previous MD simulations \cite{Saaskilahti2014,namle}.

%\underline{\sl Discussion}: 
\section{Conclusion} 

To summarize, we developed a self-consistent current-conserving approximation for anharmonic systems out of equilibrium in the high-temperature (classical) regime. There is therefore no factors of $\hbar$ in the formalism and $\om$ is to be interpreted as frequency only, not energy.
Although the set of derived equations  for the current Eq. \ref{heatcurrent} the equation of motion Eq. \ref{reom}, and the Eq. \ref{dyson} and Eq. \ref{ctot} defining the correlation functions,  were formally exact, one has to develop approximations to solve the Dyson's equation \ref{dyson} and the equation \ref{ctot} defining $C$. One, because the anharmonic force $\ar$ is an infinite Taylor expansion and is usually truncated, and two,  because its derivative appears in the denominator of Eq. \ref{dyson} which cannot be exactly inverted. In this work, we truncated the Taylor expansion of $\ar$ up to quartic terms and only included up to second powers of $\ar$ and $\partial \ar / \partial Y$ in Eqs. \ref{dyson} and \ref{ctot}. 

We showed that thermal expansion needs to be included using both cubic and quartic terms (to avoid any divergence) and it has the effect of renormalizing FCs as $T$ is increased. The reference GF to work with, $\gr$,  has two corrections: one due to thermal expansion implying changes in bond length and strength ($\phib$ instead of $\Phi$), and the other due to thermal fluctuations about the average position ($\langle \partial \ar /\partial Y \rangle$,   which usually involves the quartic term and the autocorrelation $C$), similar in spirit to the self-consistent phonon theory, except that one is not at thermal equilibrium.
%: $\gr^{-1}=[-\om^2 +\phib-\sum_{\alpha} \sigma_{\alpha} - \langle \partial \ar /\partial Y \rangle ] $ . 
This is the leading-order anharmonic correction, and cubic anharmonicity contributes to  second-order correction terms as shown in the self-consistent equations \ref{NESCP}.

Non-equilibrium averages were possible to calculate with the use of the fluctuation-dissipation theorem (Eq.  \ref{noiseautocorrelation}), the equations of motion (Eq. \ref{reom}) and the NFD theorem (Eq. \ref{NFD}). 
An alternative approach to investigate non-equilibrium effects at and near interfaces would be to perform a non-equilibrium molecular dynamics simulation (NEMD) of the system attached to thermostats at different temperatures and sample the atomic trajectories in the phase space to find the distribution functions and the position averages. This however has inherent noise in it. 

One way to extract the effective force constants is to fit from the knowledge of the forces on atoms and their positions in each MD snapshot, the forces to a linear model $F_i^{NEMD} \approx F_i^{Harmonic}=-\phib_{ij} y_j$  in order to extract the effective (non-equilibrium) harmonic force constants $\phib$. The remainder can then be defined as the anharmonic force: $F_i^{NEMD}=-\phib_{ij} y_j+a_i(y)$, and the present results maybe used.
Despite the approximations used in this work, the advantage of this formalism over MD simulations which includes anharmonicity to all orders, is that it is analytical and therefore fast and free of simulation noise, although reaching self-consistency can be challenging for some model systems. It would be desirable to make a comparison of the results with NEMD to validate these approximations for a given system.
The accuracy also
relies on the force field and strength of higher-order terms: sources of divergence would be in the denominator of Eq. \ref{dyson} if $\gr \langle \partial \ar /\partial Y \rangle \approx 1$, signaling resonances, in which case the Taylor expansion in Eq. \ref{dyson} is not appropriate.

Applications to nanoscale systems will appear in future publications.

%\underline{\sl Acknowledgments}:
\section{Acknowledgments}

We would like to thank Prof. M. R. RahimiTabar for useful discussions on the handling of correlations, the NFD theorem and a review of the manuscript, and Profs. J. Shiomi and H. Cheraghchi  for discussions at the early stages of this work. I specifically thank Dr. V. Chiloyan for introducing me to the Langevin thermostat method. Internal support at UVa from the Hobby Fund is also greatly acknowledged.

This paper is dedicated to the memory of Rouzbeh Rastgarkafshgarkolaei with whom I had several related discussions.

\appendix

\section{Calculation of time averages} \label{timeavg}

When calculating time average of a product such as $a(t) b(t)$ in terms of their Fourier transform, some care needs to be taken:
$$
\langle\langle a(t) b(t) \rangle\rangle = \frac{1}{\tau} \int_{-\tau/2}^{\tau/2} \langle a(t) b(t)\rangle \, dt ; (\tau \to \infty) $$
Where one set of brackets is for time averaging and the second set is a thermodynamic average over different initial conditions. To simplify the notations, we have however used only one set of brackets.

In terms of their Fourier transform, we have 
$$
\langle\langle a(t) b(t) \rangle\rangle = \int_{-\tau/2}^{\tau/2}
\int \frac{d\om}{2 \pi}  \frac{d\om'}{2 \pi} \,    e^{-i(\om+\om')t}\, \frac{dt}{\tau} \, \langle  A(\om) B(\om')   \rangle
$$
But 
$$ 
\langle e^{-i\om t} \rangle = 
 \int_{-\tau/2}^{\tau/2}  \, e^{-i\om t}\, \frac{dt}{\tau}  = \frac{{\rm sin} \,\om \tau/2}{\om \tau/2} \underset{\tau \to \infty}{ \to} \delta_{\om,0}  (=1 \, {\rm or} \,\, 0)
$$
On the other hand, taking $\tau \to \infty$ in the boundaries of integral of $e^{-i\om t}$  leads to $ 2 \pi \delta(\om) $ which should cancel the $\tau$ in the denominator in order to give 1.
%\begin{equation}
%\boxed{\langle e^{-i\om t} \rangle = \frac{2 \pi \delta(\om)}{\tau} = \delta_{\om,0}}    
%\end{equation}
This means, when taking time averages, one simply needs to take the convolution of the Fourier transforms at $\Om=0$:
\begin{equation}
\langle a(t) b(t) \rangle =\frac{1}{\tau} \int \frac{d\om}{2 \pi}  \, \langle  A(\om) B(-\om)   \rangle
\end{equation}
When calculating diagonal terms of autocorrelations, such as in $\langle \zeta(\om) \zeta^{\dagger}(\om) \rangle=2\gamma k_B T \tau $, the result is proportional to the integration time $\tau$. The latter cancels the $\tau $ in the denominator coming from time averaging. 
Loosely speaking,
 $2 \pi \delta(\om)/\tau =1 \, {\rm or } \, 0$.  
As an example,
to calculate the average (DC) current one simply needs to take the diagonal terms in frequencies, omitting factors of $\tau$ appearing in the correlation functions: 
\begin{widetext}
\begin{equation}
%\boxed{
\langle j_{\alpha} \rangle={\rm lim}_{\stackrel{\tau \to \infty}{\Om \to 0}} \frac{\langle J_{\alpha}(\Om)\rangle}{\tau} =\Re \int \frac{d\om}{2 \pi} \frac{(-i\om)}{\tau} \Big[\langle Y(\om) \etaad(\om) \rangle + \langle Y(\om)Y^{\dagger}(\om) \rangle \, \sa^{\dagger}(\om) \Big] 
%}
\label{heatcurrent2}    
\end{equation}
\end{widetext}

\section{Change of position variables due to thermal expansion} 
\label{xtoy}

As the temperature of a system is raised, there can the thermal expansion due to anharmonicity. The equilibrium position of the atoms is shifted, and this can also cause a change in the force constants as bond lengths have changed. To take these effects into account, we will slightly modify the formalism as follows:

Let us  Taylor-expand the interaction potential in the device in powers of rescaled atomic displacements $x={\sqrt m}\, u$, about the zero temperature equilibrium positions:
\begin{widetext}

\begin{equation}
    \V(x_1,...,x_n)=\V(0,...0)+ \Pi_i x_i +\frac{1}{2!}  \Phi_{ij} \,x_i \,x_j +\frac{1}{3!} \, \Psi_{ijk} \,x_i x_j x_k +\frac{1}{4!} \, \chi_{ijkl} \, x_i x_j x_k x_l + ... 
\label{vtot}
\end{equation}
\end{widetext}
where, for brevity, we omitted the summation sign over repeated indices, and where $\Pi_i$ is the residual force on atom $i$ and is zero if one starts from a fully relaxed configuration. The coefficients of this expansion can be obtained from a zero-temperature DFT calculation for instance\cite{esfarjani2008,Tadano2014c}.

One way, and actually the correct way to compute the non-equilibrium average of the force constants, is to perform a non-equilibrium molecular dynamics (NEMD) simulation in which the system is subject to two or more thermostats at different temperatures and zero pressure. Average positions $\langle x \rangle$ maybe computed from the runs, and then one may fit the forces with an effective harmonic model as we will describe below.  
In a thermally expanded system where $\langle x \rangle $ is non-zero, we Taylor expand the potential about the "non-equilibrium" values $\langle x \rangle$ instead of zero. This leads to a change in the force constants. 
The new dynamical variables are chosen so that their average is zero, i.e. they oscillate about the new thermal non-equilibrium positions:
$$ %\boxed{
y(t)=x(t)-\langle x \rangle %\,;\,\langle y \rangle=0
=> 
Y(\om)=X(\om)- 2 \pi \delta(\om)\langle x \rangle ; \langle y\rangle=0
$$
The average position is defined to be the solution of the average force being zero. 
 In terms of these new variables, and up to second-order in powers of displacements, the force on atom $i$ can be written as:
\begin{widetext}
\begin{equation}
    -\frac{\partial \V}{\partial x_i} = F_i = F_i(x=\langle x\rangle)+  \Big(\frac{\partial F_i}{\partial y_j}\Big)_{x=\langle x\rangle } \, y_j + \frac{1}{2} \Big(\frac{\partial^2 F_i}{\partial y_j y_k}\Big)_{x=\langle x\rangle} \, y_j y_k+ ...
\end{equation}
\end{widetext}
The first term, the residual force, will be used to define the  thermal expansion $\langle x \rangle$  (by setting the residual or average force to zero: $ F_i(x=\langle x\rangle)=-(\frac{\partial \V}{\partial y_i})_{x=\langle x\rangle} =0$), which is the Eq. \ref{favg0} in the main text. The second term is the effective, or temperature-dependent, harmonic force, and the last term is the effective anharmonic force. In the case where the potential energy 
is Taylor expanded up to quartic order such as in Eq. \ref{vtot}, one can explicitly work out the equation satisfied by $\langle x \rangle$, and derive explicit expressions for the effective harmonic and anharmonic FCs. The expression for the force in this case is as follows:
\begin{widetext}
\begin{flalign*}
    -\frac{\partial \V}{\partial x_i} &= F_i = - \Phi_{ij} x_j-\frac{1}{2} \Psi_{ijk} x_j x_k -\frac{1}{6} \chi_{ijkl} \, x_j x_k x_l  - \Sigma_{\alpha} V_{\alpha}\, x_{\alpha} \\
    %\nonumber \\
    &= - \Phi_{ij} \langle x_j \rangle -\frac{1}{2} \Psi_{ijk} \langle x_j \rangle \langle x_k \rangle -\frac{1}{6} \chi_{ijkl} \langle x_j \rangle \langle x_k \rangle \langle x_l \rangle 
    %\nonumber \\
    -\Phi_{ij} y_j-\frac{1}{2} \Psi_{ijk} (y_j y_k + y_j\langle x_k \rangle+\langle x_j \rangle y_k) \\
    %\nonumber \\
    &-\frac{1}{6} \chi_{ijkl} \,(y_j y_k y_l + 3 y_j y_k \, \langle x_l \rangle  + 3 y_j \langle x_k \rangle \langle x_l \rangle )  - \Sigma_{\alpha} V_{\alpha}\, x_{\alpha} 
   % \nonumber \\
%    &= F_i^{res} - \phit_{ij}\, y_j -\frac{1}{2} \psit_{ijk} \, y_j y_k -\frac{1}{6} \chit_{ijkl} \, y_j y_k y_l - \Sigma_{\alpha} V_{\alpha} x_{\alpha} 
%\label{force2}
\end{flalign*}
\end{widetext}

where use was made of the symmetry of $\Psi$ and $\chi$ under permutation of their $j,k,l$ indices. 

\subsection{Thermal expansion}\label{cte}

 To find the average positions $\langle x \rangle$, we set the average force on each atom $i$, $\langle F_i \rangle= - \langle \big(\partial \V / \partial x_i \big)_{x=\langle x \rangle} \rangle $ to zero,  and $\langle x_{\alpha} \rangle=0$ (leads are harmonic and will not thermal-expand), and solve for $\langle x \rangle$. Thus the coupling term with the leads which is linear in lead degrees of freedom and has zero average, will vanish, and we have the following set of non-linear equations in $\langle x \rangle$ in terms for force constants and averages of $\langle y y \rangle$ and $\langle y y y \rangle$: 
\begin{widetext}
\begin{equation} %widetext}
\Big[\Phi_{ij}\,+\frac{1}{2}\chi_{ijkl}\,\langle y_k y_l \rangle \Big]  \langle x_j \rangle + \frac{1}{2} \Psi_{ijk}\,\langle x_j \rangle \langle x_k \rangle +\frac{1}{6} \chi_{ijkl}\,\langle x_j \rangle \langle x_k \rangle \langle x_l \rangle= -\frac{1}{2} \Psi_{ijk}\, \langle y_j y_k \rangle -\frac{1}{6} \chi_{ijkl}\, \langle y_j y_k y_l \rangle 
\end{equation} %widetext}
\end{widetext}

\subsection{The new force constants}\label{newfcs} 

The new force constants are simply defined to the be potential derivatives evaluated at the new average positions. 
Up to quartic order, they are defined as:
\begin{flalign*}
\begin{alignedat}{4}
\phib_{ij} &=  \Big(\frac{\partial^2 \V}{\partial x_i \partial x_j} \Big)_{\langle x \rangle} 
= \Phi_{ij} + \Psi_{ijk}\, \langle x_k \rangle + \frac{1}{2} \chi_{ijkl}\,  \langle x_k \rangle \langle x_l \rangle 
\nonumber \\
\psib_{ijk} &= \Big( \frac{\partial^3 \V}{\partial x_i \partial x_j \partial x_k} \Big)_{\langle x \rangle} =
\Psi_{ijk}+ \chi_{ijkl} \,\langle x_l \rangle  
\nonumber \\
\chib_{ijkl}&=\chi_{ijkl}  
\end{alignedat}
\end{flalign*}

%This expansion has the advantage of incorporating most of the force in the linear term and we expect that the remaining anharmonic terms would be small.

Finally, the average positions can be re-expressed in terms of the new force constants as:
\begin{widetext}
\begin{equation} %widetext}
     %-\phib_{ij}\, \langle x_j \rangle  - \frac{1}{2} \psib_{ijk}\, \langle x_j x_k \rangle=
\boxed{ %-\frac{F_i^{res}}{\sqrt{m_i}}=
\phib_{ij}\, \langle x_j \rangle  - \frac{1}{2} \psib_{ijk}\,\langle x_j \rangle \langle x_k \rangle+\frac{1}{6} \chib_{ijkl}\,\langle x_j \rangle \langle x_k \rangle \langle x_l \rangle =-\frac{1}{2} \psib_{ijk}\, \langle y_j y_k \rangle -\frac{1}{6} \chib_{ijkl}\, \langle y_j y_k y_l \rangle  }
    \label{force0}
\end{equation} 
\end{widetext}

These coupled set of non-linear equations (for all $i$) in (${ \langle x \rangle } , \phib,\psib, \chib$ must be solved in terms of  $\langle yy \rangle$ and $\langle yyy \rangle$ covariance matrices, to find the average positions and effective force constants. It requires a self-consistent  iterative solution as the $\langle yy \rangle $ correlations will in turn depend on $ \langle x \rangle  $.

\subsection{New equations of motion}

Finally, changing to new dynamical variables $y(t)$ and new force constants, the equation of motion for $y$ becomes:
\begin{widetext}
\begin{flalign}
 \frac{d^2y_i}{dt^2}&= -\Big(\frac{\partial^2 \V}{\partial y_i \partial y_j}\Big)_{y=0} \, y_j  -\frac{1}{2}\Big(\frac{\partial^3 \V}{\partial y_i \partial y_j \partial y_k}\Big)_{y=0} \,  y_j y_k -\frac{1}{6}\Big(\frac{\partial^4 \V}{\partial y_i \partial y_j \partial y_k \partial y_l}\Big)_{y=0} \,  y_j y_k y_l - ...-\Sigma_{\alpha} V_{\alpha}\, x_{\alpha} \nonumber \\
&= 
  - \phib_{ij}\, y_j - \frac{1}{2} \psib_{ijk}\, y_j y_k  -
 \frac{1}{6} \chib_{ijkl} \, y_j y_k y_l 
- \Sigma_{\alpha} V_{\alpha}\, x_{\alpha}
  %\frac{F_i-F_i^{res}}{\sqrt{m_i}} =  - \phib_{ij}\, y_j - \frac{1}{2} \psib_{ijk}\, y_j y_k   -  \frac{1}{6} \chib_{ijkl} \, y_j y_k y_l - \Sigma_{\alpha} V_{\alpha}\, x_{\alpha}    
\end{flalign}
\end{widetext}

Adopting this change of variable, the GF keeps the same form with $\Phi$ substituted by $\phib$, and  anharmonic forces in the right-hand side of the equation of motion become by definition:
\begin{equation}
    a \myeq   - \frac{1}{2} \psib_{ijk}\, y_j y_k -\frac{1}{6} \chib_{ijkl}\,  y_j y_k y_l 
\end{equation}

%As an example, if we go up to fifth-order in the Taylor expansion of the potential energy, and use Eq. \ref{phibar} for $\bar{\phi}_{ij}$, the anharmonic force would have the following expression:
%\begin{equation}
%a    =
%    - \frac{1}{2} \Psi \Big(x^2 - 2 \langle x \rangle x \Big) -
%    \frac{1}{6} \chi \Big(x^3 - 3 \langle x^2 \rangle x \Big) -
%    \frac{1}{24} \zeta \Big(x^4 - 4 \langle x^3 \rangle x \Big) +... 
%    \label{anh}
%\end{equation}

%where the second term in each of the parentheses is the linear force included in the effective force constant: $\bar{\phi}=\phi+ \Psi \langle x \rangle+ \chi \langle x^2 \rangle/2 +  \zeta \langle x^3 \rangle/6 $ which is now subtracted from the anharmonic force.

%In general, since we will stop at cubic force constants $\psi$ in the expression of the force, for any choice of $\bar{\Phi}$ the best $\bar{\Psi}$ is found by fitting the remainder $F_i/\sqrt{m_i}- (-\bar{\Phi}_{ij} x_j) $ with $-1/2   \bar{\Psi}_{ijk} \, x_j x_k $ so that the remaining part of the anharmonic force is all lumped into $\psib$.

%The remaining constant terms in the force are: $-\phib \langle x \rangle- \psib \langle x \rangle^2 /2$ which is set to zero to find the shifted equilibrium position $\langle x \rangle$

\section{Explicit form of the correlation functions including the atomic and cartesian indices} \label{eom}

In this section, we will give explicit formulas for the correlation functions needed in the average heat current expression in Eq. \ref{heatcurrent2}. 
First let us label by roman letters $i,j,k$ the dynamical degrees of freedom in the device. If there are N atoms in the device in 3 dimensions, each of these labels refers to an atom and one of the 3 cartesian components  of its displacements. It therefore varies from 1 to 3N.
As a result the matrices $C$, $\sa$ and $\Wa$ are $3N\times 3N$ matrices. Taking their trace in formula \ref{heatcurrent2} gives the heat current, which is a scalar. Note that since transport is along the length of the leads which are one-dimensional, in principle the components of the power perpendicular to the lead direction should yield zero. More specifically, if the current in lead $\alpha$ is given by $J_{\alpha}=\langle \dot{x} F_x+\dot{y} F_y+\dot{z} F_z \rangle$, and the lead is infinite along the $z$ direction for instance, then we should have $J_{\alpha}=\langle \dot{z} F_z \rangle$ and $\langle \dot{x} F_x \rangle=\langle \dot{y} F_y \rangle=0$. So in each lead, one may take 3 partial traces along and perpendicular to the lead direction and confirm these relations. The total trace should still yield the correct result. One final note is the matrices of $\phib_{ij}$ and $\psib_{ijk}$ are invariant under permutations of their indices.

\subsection{Lead Self-energies $\sa$ and escape rates $\Ga$}

The surface green functions of the leads were defined by an inverse, as stated in Eq. \ref{sgf}. The self-energy for lead $\alpha$ is accordingly defined as
$\sigma_{\alpha,ij}(\omega)= V_{\alpha,ik} \, \gamma_{\alpha,kk'}(\omega) \, V_{\alpha,jk'}$ where the two indices $(k,k')$ refer to the lead degrees of freedom. Note the matrices $\Va$ connecting the device to lead $\alpha$ need not be square. 

The escape rates were defined by:
\begin{equation}
    \Gamma_{\alpha,ij} =-i(\sigma_{\alpha,ij}-\sigma_{\alpha,ij}^{\dagger})=2 \Im(\sigma_{\alpha,ij})
\end{equation}
For Langevin thermostats with white noise, these results do not depend on the thermostat damping factors $\gamma$.
%These results depend on the adopted damping factor for the leads $\gama$, but in the limit $\gama \to 0$ they tend to unique well-defined matrices which we will use for the escape rate matrices. The strength of the latter can be controlled with non-zero $\gama$ or stronger coupling $\Va$ matrices. 

\subsection{Noise autocorrelation functions}
\label{noiseFD}

Let us start with the noise autocorrelation which will appear in the calculation of displacement-noise correlation $\Wa$. This autocorrelation can be obtained from the fluctuation-dissipation theorem, a relation which has to hold if the noise is such that the leads are to behave as a Langevin thermostat at temperature $T$:
$$\langle \zeta_{\alpha}(t) \zeta_{\alpha'}^T(t') \rangle \stackrel{FD}{=} 2 \gama k_B T_{\alpha} \,\delta(t-t')\delta_{\alpha,\alpha'}$$
The noise is white and different sites are uncorrelated with each other. 
This relation implies that in the frequency domain the autocorrelation of $\zeta$ and that of  $\eta$ satisfy the following relations:
\begin{equation}
    \langle \zeta_{\alpha}(\omega) \zeta_{\alpha'}^T(\omega') \rangle = 2\pi \delta(\omega+\omega')\,\delta_{\alpha,\alpha'} \, 2 \gama   k_B T_{\alpha}
\end{equation}
%(The quantum version is:
%\begin{equation}
%    \frac{1}{2} [\langle \zeta_{\alpha}(\omega) \zeta_{\alpha'}^T(\omega') \rangle+\langle \zeta_{\alpha}(\omega') \zeta_{\alpha'}^T(\omega) \rangle] 
%    =  \delta(\omega+\omega')\,\delta_{\alpha,\alpha'} \,  \gama \omega {\rm coth}\, \frac{\hbar \omega}{2 k_B T_{\alpha}}  
%\end{equation}
%or
%\begin{equation}
%    \langle \zeta_{\alpha}(\omega) \zeta_{\alpha'}^T(\omega') \rangle = 2\pi \delta(\omega+\omega')\,\delta_{\alpha,\alpha'} \, 2 \gama \omega [1+f_{BE}(\frac{\hbar \omega}{ k_B T_{\alpha}})]  
%\end{equation}
%)

\begin{flalign}
    &\langle \eta_{\alpha}(\omega) \,\eta_{\alpha'}^T(\omega') \rangle = \Va  \ga \
    ,  \langle \za \zeta_{\alpha'}^T \rangle \,g_{\alpha'}^T V_{\alpha'}^T \\&=2\pi \delta(\omega+\omega')\delta_{\alpha,\alpha'}\, k_B T_{\alpha}\, [\Va  \ga \, (2 \gama)  \,g_{\alpha}^T V_{\alpha}^T]   
\end{flalign}
In the above, $\ga$ is a diagonal matrix of size equal to the number of degrees of freedom in the lead $\alpha$ (which is infinity!) 
At this point, we will  use a convenient identity satisfied by any Green's function $G$. If $G^{-1}=a+ib$, with $(a,b)$ real matrices, then
\begin{equation}
    i(G-\Gd)=iG(\frac{1}{\Gd} - \frac{1}{G})\Gd= %iG[(a-ib) -(a+ib)]\Gd=
    G \, (2b) \, \Gd %= G \, (2 \Im G^{-1}) \, \Gd
    \label{greenid}
\end{equation}
Since $\Im (g_{\alpha}^{-1})=-\om \gama$  we can write the $\eta$-autocorrelation as:
\begin{widetext}
\begin{flalign}
        \langle &\eta_{\alpha}(\omega) \,\eta_{\alpha'}^T(\omega') \rangle  %\Va\ga \,(2\omega \gama)\ga^{\dagger} \Va^T  2\pi \delta(\omega+\omega')\delta_{\alpha,\alpha'}\, \frac{k_BT_{\alpha}}{\omega} \nonumber\\ 
        = -i\Va(\ga-\ga^{\dagger})\Va^T \times 2\pi \delta(\omega+\omega')\delta_{\alpha,\alpha'}\, \frac{k_BT_{\alpha}}{\omega} \nonumber\\
        &= -i(\sa-\sa^{\dagger}) \times 2\pi \delta(\omega+\omega')\delta_{\alpha,\alpha'}\, \frac{k_BT_{\alpha}}{\omega} %\nonumber\\ &
        = \Gamma_{\alpha} \times 2\pi \delta(\omega+\omega')\delta_{\alpha,\alpha'} \, \frac{k_BT_{\alpha}}{\omega}
        \label{noiseautocorrelation2}
\end{flalign}
\end{widetext}
 where we used  the notation $\Gamma_{\alpha} =-i(\sa-\sa^{\dagger})=2 \Im (\sa) $ for twice the imaginary part of the lead $\alpha$ self-energy. 
 Alternatively, the diagonal elements in frequency can be written as:
 \begin{flalign}
\langle \eta_{\alpha}(\omega) \eta_{\alpha'}^T(-\omega) \rangle =
\Gamma_{\alpha} \times  \delta_{\alpha,\alpha'} \, \frac{k_BT_{\alpha}}{\omega} \tau 
 \end{flalign}
 where $\tau$ is the integration time which goes to infinity.
 Dhar and Roy\cite{Dhar2006} have shown that in the quantum limit, when taking a semi-infinite harmonic lead and averaging over all possible initial conditions sampled from a canonical ensemble, the quantum noise term has an autocorrelation given by $$\langle \eta_{\alpha}(\omega) \eta_{\alpha'}^T(\omega') \rangle=  h \Gamma_{\alpha}  \delta(\omega+\omega')\delta_{\alpha,\alpha'} \Big(1+f(\hbar\omega/k_BT)\Big) $$ where $f(x)=[e^x-1]^{-1}$ is the equilibrium Bose-Einstein distribution function, which, in the classical (high-temperature) limit, reduces to $k_BT/\hbar \omega$. Our classical result based on properties of Langevin thermostats is thus consistent with the quantum one.
 %and to extend to the quantum case, we can  substitute  $ (k_BT/ \omega)$ by $\hbar (1+f(\hbar \omega/k_BT)) $. 
 
 We can note that the explicit dependence on the damping factor $\gama$ has been replaced by twice the imaginary part of the lead self-energy $\Ga$, which one may interpret as $2 \om$ times a ``rate''. 
 %The thermostat parameter $\gamma$ can now be taken to zero, so that the results do not artifically depend on it. %provided we take $\omega \gama=\Ga$.
 For the adopted white noise, the dependence on its damping factor has gone away!
 Furthermore, in the calculation of  $\Wa$, the notation $\langle \eta \eta^{\dagger} \rangle$ implies the diagonal terms $\om'=-\om$ must be taken. Using the Novikov-Furutsu-Donsker (NFD) identity (see appendix \ref{nfdapp}) we can see that $\langle A \eta^{\dagger} \rangle \propto \langle \eta \eta^{\dagger} \rangle $ and therefore only diagonal terms in frequency will appear in the frequency integral of the heat current, and assuming thermostat temperatures are steady, the heat current can only have a DC ($\Om=0$) component. Note that this component is infinite the way we defined it: $\Ja (\Om=0) = \int_{-\infty}^{\infty} \langle j_{\alpha}(t) \rangle \, dt$. To find the average heat current, we have to divide this by the integration time $\tau$, and then take the limit $\tau \to \infty$. This division cancels  the $\tau$ appearing in the numerator of noise autocorrelations. This issue is further discussed in the appendix, Eq. \ref{heatcurrent2}. Final results do not depend on $\tau$.

 The factor  $2 \pi \delta(\omega+\omega') /\tau $ can now be excluded from the current as it is essentially 1 for the diagonal terms, and zero otherwise. 
 \begin{equation}
\boxed{ \langle \eta_{\alpha}(\om) \eta_{\alpha}^{\dagger}(\om) \rangle = \Ga(\om) \frac{k_BT_{\alpha}}{\omega} \tau=2 \Im(\sa(\om))  \frac{k_BT_{\alpha}}{\omega} \tau  } 
 \end{equation}

\subsection{Noise-displacement correlations $\Wa$} \label{zalpha}

Using the equation of motion \ref{reom}, the expression defining this correlation function to second order in $\partial \ar/\partial Y$ can be derived to satisfy Eq. \ref{2nd_order_z}. After substitution, the explicit relations become:
\begin{widetext}
\begin{flalign}
&Z_{\alpha,ij}(\om) = Z^{NEMF}_{\alpha,ij}(\om)+ 
 \gr_{ik}(\omega) \,\Big(\Sigma^{(3)}_{kk'}+\Sigma^{(4)}_{kk'} \Big)\,  Z_{\alpha,k'j}(\omega) \nonumber \\  
 &Z^{NEMF}_{\alpha,ij}(\om) = \gr_{ik}(\om) \, \Gamma_{\alpha,kj}(\om) \frac{k_B \Ta}{\om} \nonumber \\
&\Sigma^{(3)}_{kk'}(\om) = 2 \psib_{klm}\, \psib_{k'l'm'} \Big(\int_1 \gr_{ll'}(\om-\omega_1) \, C_{mm'}(\omega_1) \Big) \nonumber \\
&\Sigma^{(4)}_{kk'}(\om) = \frac{3}{2} \chib_{klmn} \, \chib_{k'l'm'n'} \Big( \int_{1,2} \gr_{ll'}(\om-\om_1-\om_2) \, C_{mm'}(\omega_1) \,C_{nn'}(\omega_2) \Big) 
\label{zeqs} 
\end{flalign}\end{widetext}

where, as usual, the summation over repeated indices is implied.

\begin{figure}[!h]
\centering
\includegraphics[scale=0.40]{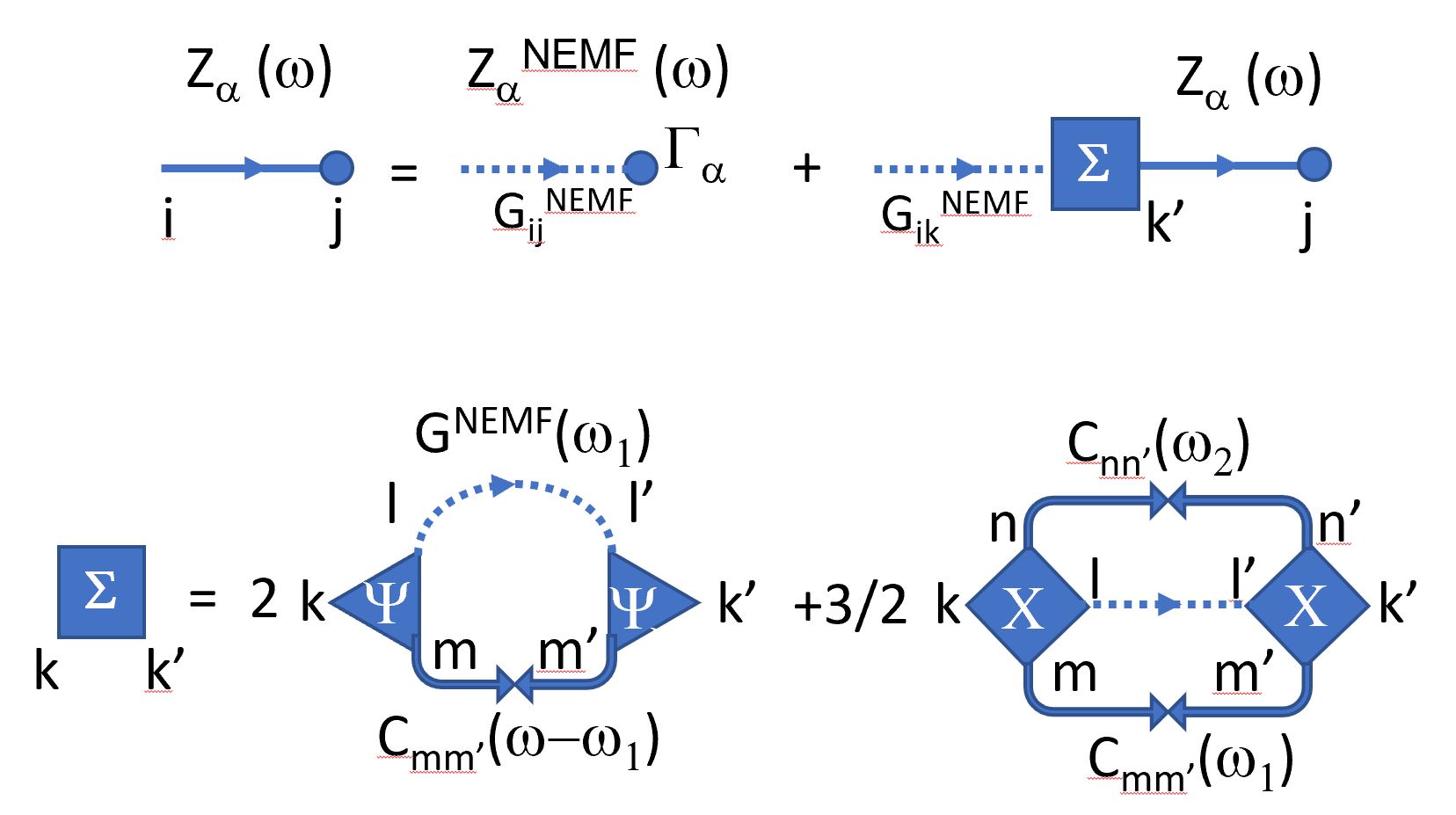} %{Z34.png}
\caption{Feynman diagrams associated with $\Wa$ up to second order in anharmonic forces. Dashed lines represent the phonon Green's function $\gr$, the circle represents $\Ga$,
the triangle represents the third-order vertex $\psib$ and the square the fourth-order vertex $\chi$. The thick line with opposite arrows represents the displacement autocorrelation  $C(\om)$. Frequency must be conserved at each vertex.
}
\label{zfeynman}
\end{figure}

\subsection{Displacement autocorrelations $C(\om)$}\label{yy}

These functions, whose trace represents the number of excitations in the device, are defined via Eq \ref{ctot}. The static distortion in $\langle x\rangle  \langle x\rangle $ does not contribute to the current and will thus be omitted. To second-order in $\ar$, the expression for $C$ is given by:
\begin{widetext}
\begin{flalign}
&C_{ij}(\om) =\sum_{\alpha} (Z_{\alpha,ik}-Z^{NEMF}_{\alpha,ik}) \gr_{kj}^{\dagger} + \gr_{ik}  (Z_{\alpha,kj}-Z^{NEMF}_{\alpha,kj})^{\dagger} - C_{ij}^{NEMF}(\omega) + \gr_{ik} \,\Big(C^{(3)}_{kk'}+C^{(4)}_{kk'} \Big)\,  \gr_{k'j}^{\dagger} \nonumber \\  
&  C_{ij}^{NEMF}(\om)=  \Sigma_{\alpha} \, (\gr_{ik} \, \Gamma_{\alpha,kk'} \, \gr^{\dagger}_{k'j} ) \,\frac{k_BT_\alpha }{\omega} \nonumber \\
&C^{(3)}_{kk'}(\om) = \frac{1}{2} \psib_{klm}\, \psib_{k'l'm'} \Big(\int_1 C_{ll'}(\om-\omega_1) \, C_{mm'}(\omega_1) \Big) \nonumber \\
& C^{(4)}_{kk'}(\om) = \frac{1}{6} \chib_{klmn} \, \chib_{k'l'm'n'} \Big( \int_{1,2} C_{ll'}(\om-\om_1-\om_2) \, C_{mm'}(\omega_1) \,C_{nn'}(\omega_2) \Big)
\label{ceqs}
\end{flalign}
\end{widetext}

\begin{figure}[!h]
\centering
\includegraphics[scale=0.35]{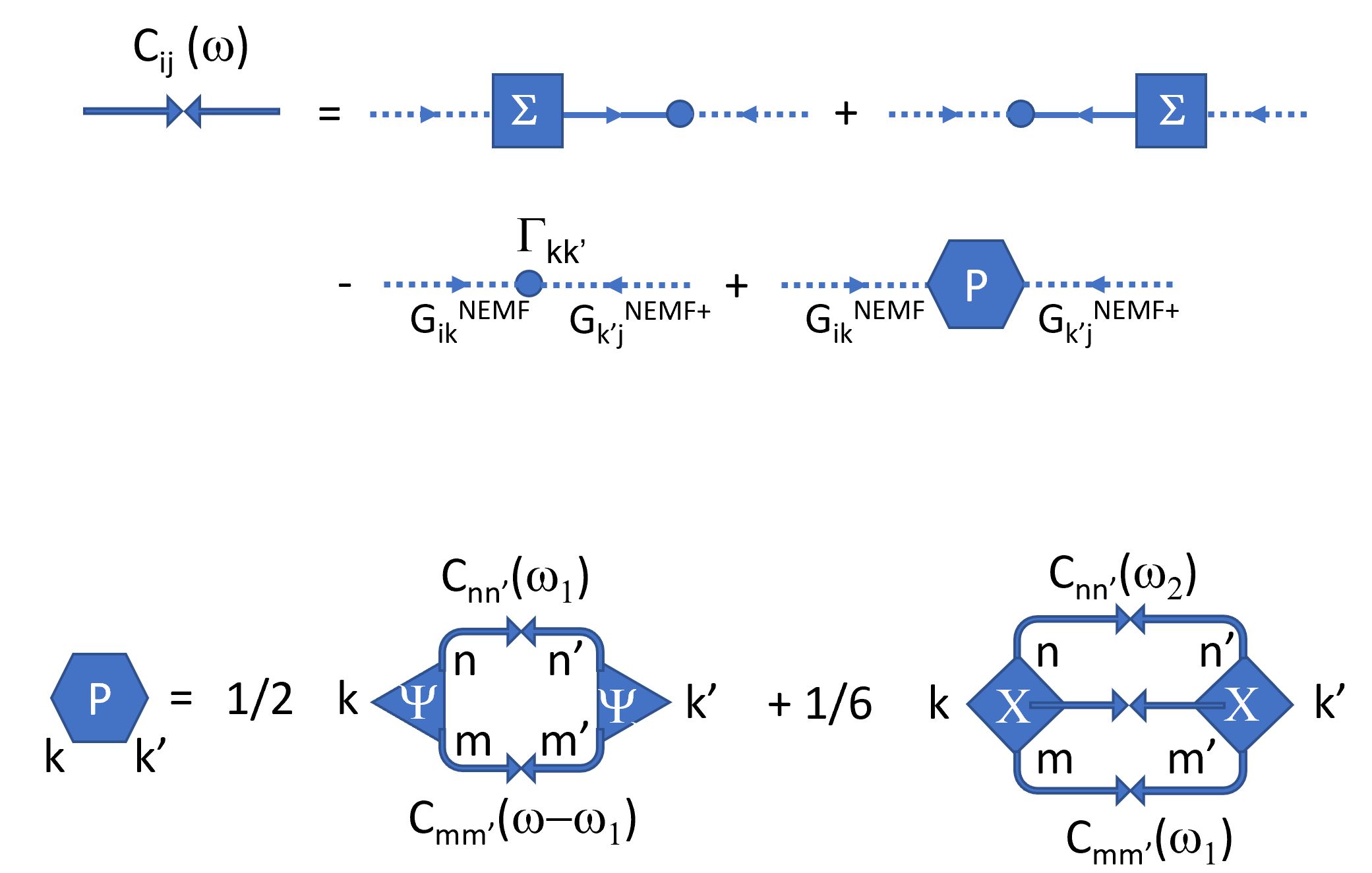} %{C34.png}
\caption{Feynman diagrams associated with $C(\om)$ up to second order in anharmonic forces. Conventions are the same as in Fig. \ref{zfeynman}
}
\label{cfeynman}
\end{figure}

 \section{Statement and proof of the Novikov-Furutsu-Donsker (NFD) relation} \label{nfdapp}
 
The Novikov-Furutsu-Donsker relation\cite{donsker1962,furutsu1964,novikov1965,tabar2019analysis} relates the correlation function of any functional $A[\eta]$ with the noise $\eta$ to the noise-noise correlation times the expectation value of the derivative of that functional:
\begin{equation}
\langle A[\eta] \, \eta_{\alpha}^{\dagger}(t) \rangle= \sum_{\beta} \int dt' \langle \eta_{\beta}(t')  \eta_{\alpha}^{\dagger}(t) \rangle \langle \frac{\delta A[\eta]}{\delta \eta_{\beta}(t')} \rangle    
\end{equation}
implying that in the frequency domain, we have
\begin{equation}
\langle A_i[\eta] \, \eta_{j\alpha}^{\dagger}(\om) \rangle  =  \sum_{l,\beta} \langle \frac{\delta A_i[\eta]}{\delta \eta_{l,\beta}}(\om) \rangle 
\langle \eta_{l,\beta}(\om)  \eta_{i,\alpha}^{\dagger}(\om) \rangle
%\Ga_{,lj} \frac{k_B T_{\alpha}}{\om} \tau=   \langle \frac{\partial A_i}{\partial Y_k} \rangle \, \gf_{kl} (\om)\, \Ga_{,lj} \frac{k_B T_{\alpha}}{\om} \tau
\end{equation}

{\it Proof of the NFD relation}: Let $\eta $ be a random variable with Gaussian distribution of mean 0 and variance $\sigma^2$.
Consider the average $S=\langle A[\eta] \eta \rangle$ where $A$ is a functional of $\eta$. By definition,  $S=\int_{-\infty}^{+\infty} d\eta\frac{e^{-\eta^2 /2 \sigma^2}}{\sqrt{ 2 \pi \sigma^2}}   \eta  A(\eta)$.
After performing an integration by parts, we find:
\begin{widetext}

$$S= \int_{-\infty}^{+\infty} -\sigma^2 \frac{1}{\sqrt{ 2 \pi \sigma^2}} d(e^{-\eta^2 /2 \sigma^2}) A(\eta)=\sigma^2 \int_{-\infty}^{+\infty} \frac{1}{\sqrt{ 2 \pi \sigma^2}} e^{-\eta^2 /2 \sigma^2} \frac{d A(\eta)}{d\eta} d\eta= \sigma^2 \langle \frac{d A(\eta)}{d\eta} \rangle =   \langle \eta \eta \rangle  \langle \frac{d A(\eta)}{d\eta} \rangle 
$$
\end{widetext}

This can easily be extended to higher dimensions where $\eta$ is an array: $\langle A[\eta] \eta_i \rangle= \sum_j \langle \frac{\partial A}{\partial \eta_j} \rangle \langle \eta_j \eta_i \rangle$.

\section{Heat current within the Harmonic approximation} \label{harmonic}

For a harmonic system ($G=\gf=\gr$), we have derived the following expressions :
\begin{flalign}
\Wa^H(\om)&=\langle  Y \eta^{\dagger}_{\alpha}\rangle =\gf \Ga \frac{k_BT_{\alpha}}{\om} \tau; \\ C^H(\om)&=\langle  Y Y^{\dagger}\rangle=\Sigma_{\beta} (\gf \Gamma_{\beta} \gf^{\dagger})   \frac{k_BT_{\beta}}{\om} \tau  
\end{flalign}

The frequency in the denominator will cancel the frequency in the current coming from the time derivative of positions, so that the harmonic current becomes:
\begin{equation}
\langle  j_{\alpha}^H \rangle =k_B \int \frac{d\omega}{2 \pi} \,  {\rm Tr}\, \Big(  \Im (\gf) \,\Ga  T_{\alpha} - 
\Sigma_{\beta} (\gf \,\Gamma_{\beta} T_{\beta}\,\gf^{\dagger}) \frac{\Ga}{2}     \,   \Big)     \nonumber
\end{equation}

Below we will show that this current satisfies both detailed balance and current conservation. 

{\it Detailed balance}: If all leads are at the same temperature, using Eq. \ref{greenid}, we have
%the currents should identically be zero:
\begin{equation}
  {\rm Tr}\, \Ga \,  \Im (\gf)  = \frac{1}{2} {\rm Tr}\, \Ga
\Sigma_{\beta} (\gf \Gamma_{\beta} \gf^{\dagger})         
\end{equation}
Making this substitution for $\Im (\gf)$ in the heat current formula, we end up with:
\begin{equation}
\langle   j_{\alpha}^H \rangle =\frac{k_B}{2}  \int \frac{d\omega}{2 \pi} \,  {\rm Tr}\, \Ga \,\gf \,\Big(  
\Sigma_{\beta}  \Gamma_{\beta}(  T_{\alpha}-T_{\beta})\Big) \gf^{\dagger}            
\end{equation}
This is equation manifestly shows detailed balance as it is linear in temperature differences. It is also easy to see that this harmonic part of the heat current satisfies current conservation, due to the antisymmetric (under exchange of $\alpha$ and $\beta$) form of the following sum: 
\begin{widetext}
\begin{equation}
\Sigma_{\alpha} \langle   j_{\alpha}^H \rangle =\frac{k_B}{2}  \int \frac{d\omega}{2 \pi} \, \Sigma_{\alpha\beta}\, (T_{\alpha}-T_{\beta})\, {\rm Tr}\, \Big[  \Ga \,\gf \,   \Gamma_{\beta} \gf^{\dagger}  \Big] =0 %\nonumber
\end{equation} 
\end{widetext}

In a two-terminal device geometry ($\beta=R ; \alpha=L$), the expression for the current reduces to the well-known formula:
\begin{equation}
\langle  j_{L}^H \rangle =\frac{k_B}{2} (  T_{L}-T_{R}) \int \frac{d\omega}{2 \pi} \,  {\rm Tr}\, [\Gamma_L \,\gf \,  
 \Gamma_{R}\, \gf^{\dagger}  ]     =-  \langle j_{R}^H \rangle    
\end{equation}
In general, ${\rm Tr}\, [\Ga \,\gf \, \Gamma_{\beta}\, \gf^{\dagger}  ] $ maybe interpreted as the harmonic transmission from lead $\alpha$ to lead $\beta$.
We see that even in the non-equilibrium regime (large $\Delta T$), the harmonic approximation leads to the same transmission function regardless of how large the temperature difference is.

%\subsection
{\noindent \it A simple extension to quantum case}:\\
Given the correspondence between the quantum and classical versions of noise autocorrelation, to recover the quantum limit, one may replace $f=k_BT/\omega$ by $\hbar(1+f_{\rm BE}(\hbar\om/k_BT))$. The constant term 1 is irrelevant and disappears due to the principle of detailed balance, and the temperature difference is replaced by the difference in the distribution functions times the phonon energy:$\, k_B (T_{\alpha}-T_{\beta}) <=> \hbar \om (f_{\alpha}-f_{\beta})$.

\bibliographystyle{apsrev4-1} %aipauth4-1}
\bibliography{references}

\begin{thebibliography}{27}
\bibitem [{hba()}]{hbar}%
  \BibitemOpen
  \href@noop {} {\bibinfo  {journal} {With a factor of $\hbar$ in the
  denominator $f_{\alpha}$ would be a real Bose-Einstein occupation factor
  taken to the classical limit}\ }\BibitemShut {NoStop}%
\end{thebibliography}
\end{document}